\documentclass{article}
\usepackage[utf8]{inputenc}
\usepackage{physics,graphicx,color,csquotes}
\usepackage{amsmath}
\usepackage[english]{babel}
\usepackage{caption}

\newcommand{\HBT}{Hanbury Brown and Twiss }

\begin{document}

\title{Two particle interference: The Hanbury Brown Twiss and Hong Ou Mandel experiments}
\author{Christoph I. Westbrook \\
Université Paris-Saclay, Institut d’Optique Graduate School, CNRS, \\ Laboratoire Charles Fabry, 91127, Palaiseau, France}

\date{October 2026}

\maketitle

\begin{abstract}
\noindent 
I discuss two important quantum optics experiments which depend on two particle interference.
In the years since their first demonstration many interesting and instructive variants have appeared, including the use of atoms instead of photons as the interfering particles. 
This chapter gives a pedagogical overview of some of these results and argues that the two effects are closely related.
\end{abstract}

\section{Introduction}
\begin{figure}[htbp] 
   \centering
   \includegraphics[width=7cm]{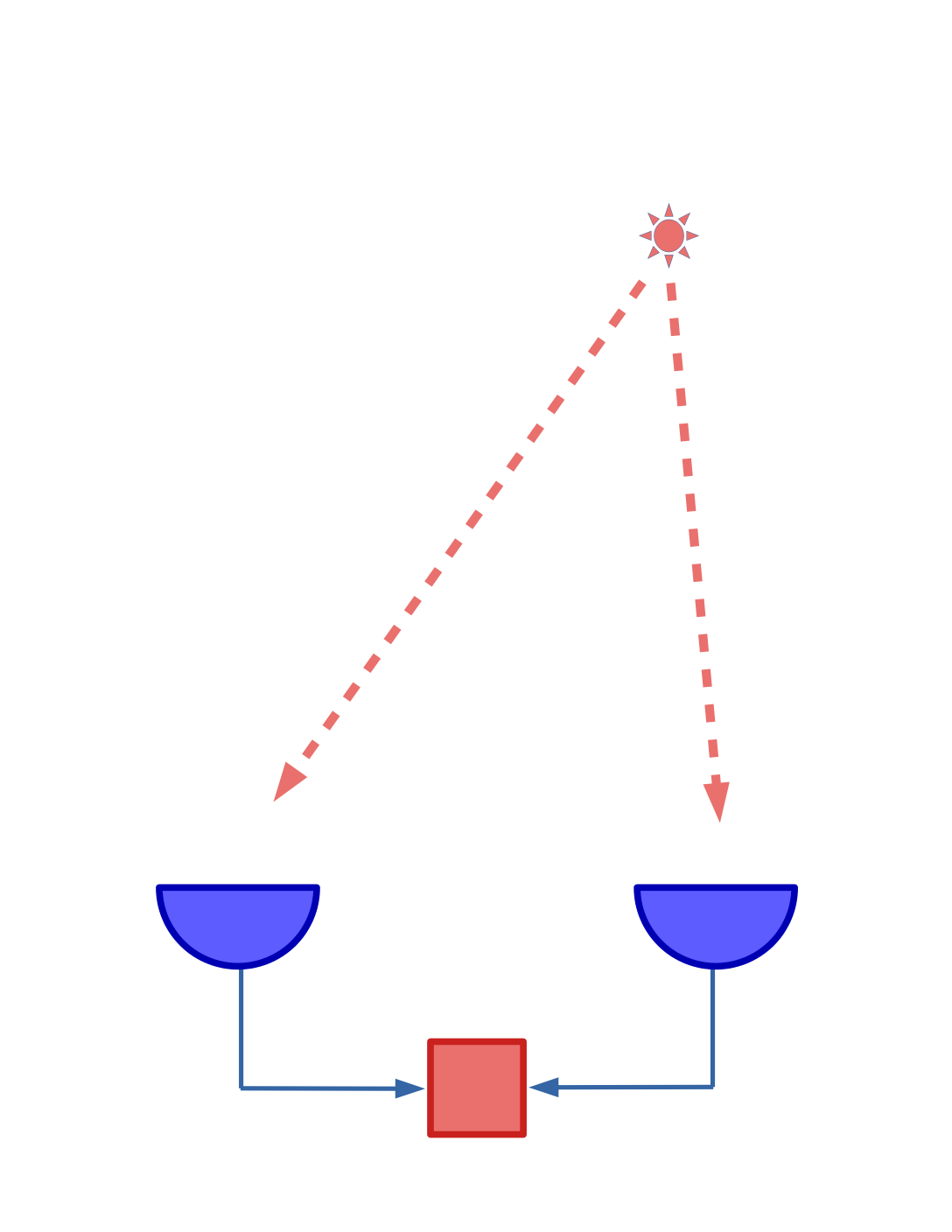} 
   \caption{ 
  Essential elements of an intensity interferometer. 
   Two receivers shown as semi-circles, convert incoming radiation from a distant object into electrical currents and an electronic circuit, shown as a red square, multiplies the two currents and averages them over some time. 
   At small detector separations, the current fluctuations are correlated, but this correlation decays with increasing separation of the detectors. 
   The characteristic distance of this decay allows one to measure the angular size of the source.
 } 
   \label{f:HBTConcept}
\end{figure}

In the 1950's, \HBT suggested a method to determine the angular size of stars using the correlation between the fluctuations received at widely separated telescopes \cite{hanbury_brown_apparent_1952, hanbury_brown_new_1954}. 
The method was first imagined for radio waves, but later it was generalized to the optical domain \cite{hanbury_brown_test_1956, hanbury_brown_intensity_1974}.
The method also continues to be of interest to the astronomy community (see for example Ref.~\cite{Guerin_Stellar_2025})
The basic apparatus is conceptually simple and is shown in Fig.~\ref{f:HBTConcept}. 
One simply uses receivers at two positions and converts the (fluctuating) electromagnetic field intensity into an electrical current. 
One then electronically multiplies the signals from the two receivers and averages this product over time.
\HBT showed that this product takes some value for zero separation and that it decays to a lower one with the distance between the two receivers. 
The characteristic decay distance is inversely proportional to the angular size of the star.  

When the method was first proposed  for the optical domain, it was met with some skepticism.
The receivers were photomultipliers which depended on the photoelectric effect, a process that was understood in terms of photons, or light quanta.
The photo current thus consisted of short pulses (or ``clicks") corresponding to the photon arrival time. 
But when thinking in terms of photons it is not at all obvious how two photons, emitted from different points on a source as incoherent as a star should conspire to arrive at separated detectors on earth in a correlated way.
Although in most cases the effect can be explained without appealing to quantum mechanics, the proposal and its success stimulated deep questions about the quantum description of light and in some sense gave rise to the field of quantum optics. 
Later experiments, notably that of Hong Ou and Mandel used a similar configuration but the results are even less intuitive and do require the quantum theory for a full explanation.
The remainder of this chapter will describe some facets of these effects and how they are related to each other.
It will also describe some generalizations to particles other than photons.

\section{Classical description}
\label{sec:classical}
\begin{figure}[htbp]
\begin{center}
   \includegraphics[width=10cm]{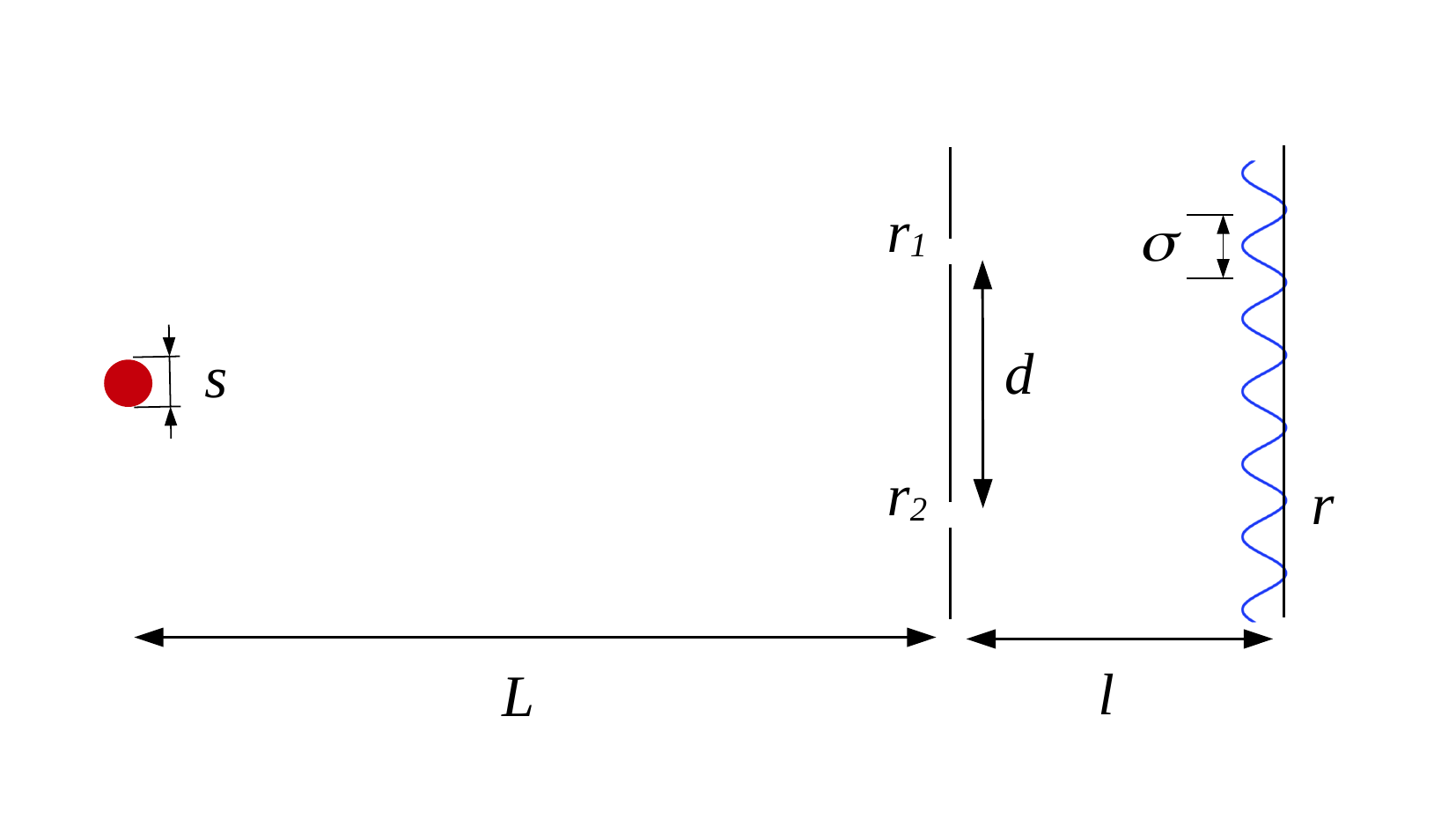} 
\caption{Basic configuration of a stellar interferometer. 
A source of size $s$ emits light which passes through two slits. 
A sufficiently small source produces fringes at the detection plane.
The maximum distance $d$ for which fringes are still visible yields an estimate for the source size. 
In the Michelson and Pease experiment the slits corresponded to mirrors which reflected the light towards a telescope mirror which in turn focused the light on an eyepiece.}
\label{f:YoungsSlits}
\end{center}
\end{figure}

Before describing the quantum theory of this type of measurement, we will first give an entirely classical description. 
We will show that for the purposes astronomical observations, a classical description is sufficient. 
Since it is closely related to the \HBT (HBT) interferometer, we will first briefly discuss a more traditional method to determine angular sizes, often referred to as stellar interferometry \cite{monnier_optical_2003, glindemann_principles_2011}. 
Stellar interferometry was suggested as early as 1868 by Fizeau \cite{fizeau_prix_1868} and first successfully carried out by Michelson and Pease \cite{michelson_measurement_1921} in 1921. 
The interferometer is essentially a Young's double slit apparatus as shown in Fig.~\ref{f:YoungsSlits}. 
There is a distant source of size $s$ whose light passes through two slits separated by a distance $d$ before arriving at position $r$ at an observation plane. 
The problem is to determine the angular size $s/L$ from the interference pattern.
The figure is not to scale and we will assume $L \gg s \gg l \gg d$ so that all propagation is paraxial. 

If the source is a monochromatic point source with wavelength $\lambda$, fringes with a spacing $\sigma =(\lambda / d) l$ will appear at the viewing plane. 
If the source has a non-zero size and if all points in the source have random relative phases, we can think of each point in the source as producing a slighty shifted fringe pattern. 
Two source points separated by distance $s$ will produce fringes shifted by $ (s/L) l$.
If one sums over all the points in the source, the fringes at the detection plane will be washed out if the typical shift is larger than the fringe separation, that is if $s/L > \lambda / d$.
Thus one can determine $s/L$ by knowing $\lambda$ and measuring the distance $d$ at which the fringes begin to wash out. 

This idea can be made more precise by considering the electric field observed at the detection plane.
(This discussion follows that of Ref.~\cite{glauber_optical_1965})
Let the electric field at the plane of the slits be given by:
\begin{equation}
E= \frac{1}{2} (E^{+}(\mathbf{r}_\mathrm{slit}) e^{-i \omega t} + E^-(\mathbf{r}_\mathrm{slit})e^{i \omega t})
\end{equation}
Here $\mathbf{r}_\mathrm{slit}$ is a position at the plane containing the slits and $\omega= 2 \pi c / \lambda$. We suppose that $E^{+}=  (E^-)^*$ is a complex quantity with a possibly fluctuating phase.
The vector nature of the field plays no important role so we will neglect it in this discussion.
The field at the viewing screen is given by
\begin{equation*}
E^{+}(\mathbf{r},t)
= K_1(\mathbf{r},t,\mathbf{r}_1,t_1) E^{+}(\mathbf{r}_1,t_1) 
+ K_2(\mathbf{r},t,\mathbf{r}_2,t_2) E^{+}(\mathbf{r}_2,t_2)
\end{equation*}
Where $\mathbf{r}$ is a point on the observation plane, $\mathbf{r}_1$ and $\mathbf{r}_2$ are the locations of the two slits, and $t_1$ and $t_2$ are the propagation times from the source to the slits.
The quantities $K_1$ and $K_2$ describe the propagation of the wave from each slit to the point $\mathbf{r}$. 
The intensity at the observation plane is 
$I= c\, \epsilon_0 \langle E^2 \rangle$ 
and is to be understood as a power per unit area.
In this section, the $\langle ... \rangle$ brackets denote an average over a time long compared to any fluctuations.
It can be written
\begin{align}
I /(\, c\, \epsilon_0)=& \,  \abs{\langle E^{+}(\mathbf{r},t) \rangle}^2\\
=&\,\abs{K_1 \langle E^{+}(\mathbf{r}_1,t_1)\rangle }^2
+\abs{K_2 \langle E^{+}(\mathbf{r}_2,t_2) \rangle }^2\\
&+\mathrm{Re}(K_1 K_2^* \langle  E^{-}(\mathbf{r}_1,t_1)  E^{+}(\mathbf{r}_2,t_2) \rangle)
\end{align}

The last term is responsible for the interference pattern shown in Fig.~\ref{f:YoungsSlits}.
The important point is that the interference term is proportional to the correlation function of the  electric field:
\begin{equation}
G^{(1)}=\langle  E^{-}(\mathbf{r}_1,t_1)  E^{+}(\mathbf{r}_2,t_2) \rangle.
\label{e:classical_g1}
\end{equation}
This function is usually referred to as the 1st order correlation function. 
We will assume that the fields are ``stationary" meaning that mean values such as $\abs{\langle E^{+}(\mathbf{r}_1,t_1)\rangle }^2$  are independent of time.
Thus in the above expression only the function $G^{(1)}$ is time dependent, and it only depends on the time difference $\tau = t_1-t_2$.
If the source emits a temporally fluctuating electric field, the correlation vanishes if the difference in propagation time from the slits to the observation plane is too long. 
According to the Wiener-Khintchin theorem \cite{loudon_quantum_2000}, for a thermal source, the timescale of the fluctuations is the inverse of the frequency bandwidth, that is for a star, femtoseconds. 
The large bandwidth also washes out the fringes far from the symmetry axis, but at the center, at least one fringe can be seen if the source is small enough.
Indeed the envelope of the fringe pattern is given by the Fourier transform of the source distribution: this is a particular case of the Van Cittert Zernike theorem in optics \cite{born_principles_1959}.

Observing this interference with starlight is difficult because other effects also wash out the interference pattern.
In particular, the positions of the slits fluctuate and atmospheric turbulence disturbs the optical paths. 
In their observations, Michelson and Pease achieved effective slit separations $d$ of up to 6 m and were thus able to measure angular diameters on the order of $10^{-7}$ (about 20 milli arc sec).
Today, with modern technology, especially adaptive optics, one can at least partially compensate for the parasitic fluctuations and thus achieve much larger separations \cite{monnier_optical_2003,glindemann_principles_2011}. 

\begin{figure}[htbp] 
   \centering
   \includegraphics[width=10cm]{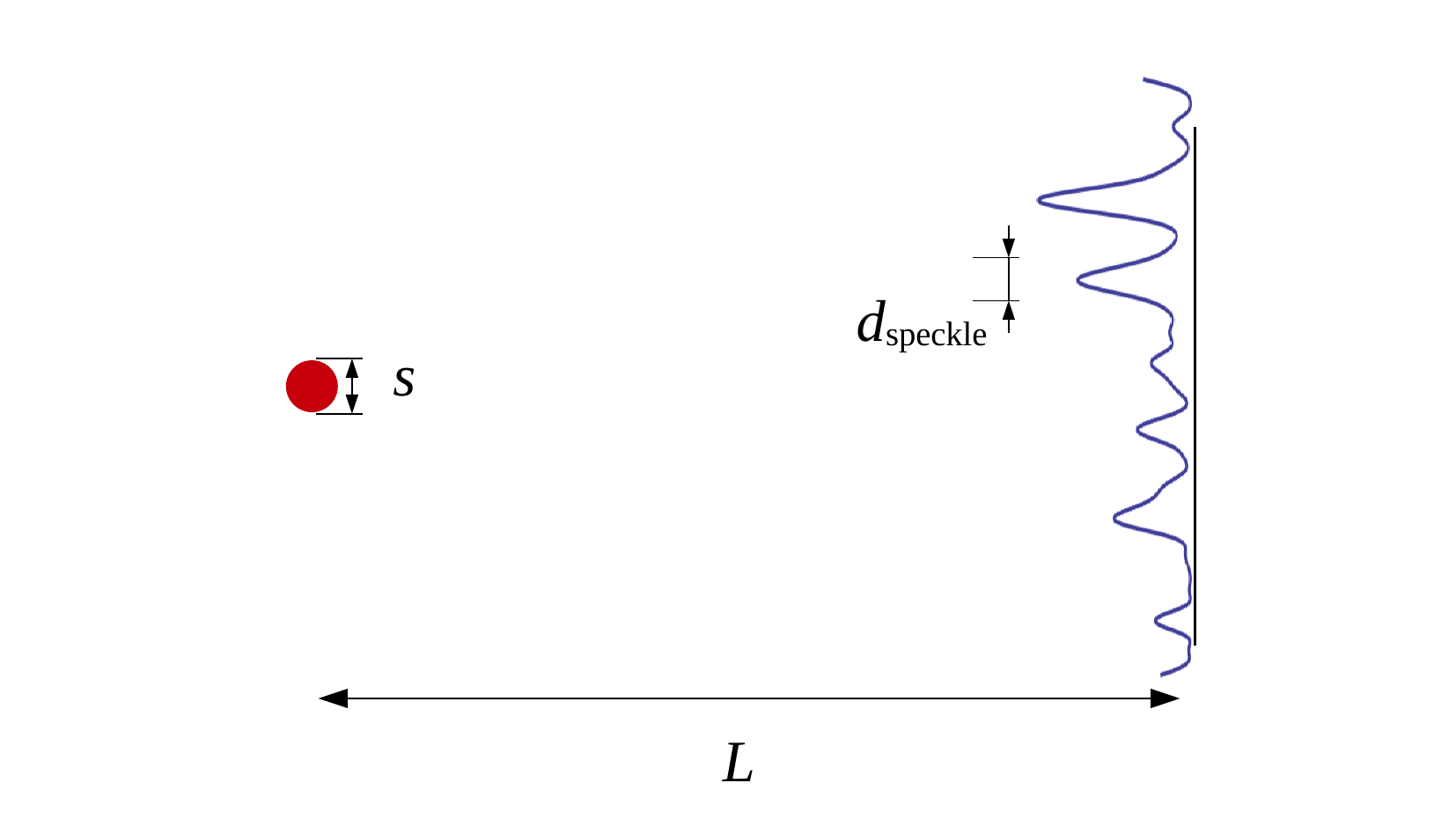} 
   \caption{
   Diagram showing the speckle at a detection plane produced by an extended light source of size $s$. 
   The characteristic size of the speckles allows one to measure the angular size of the source.} 
   \label{fig:Speckle}
\end{figure}

The \HBT method offers another way to circumvent the problem of parasitic fluctuations. 
Instead of using slits, one simply observes the intensity at different points on the observation plane as shown in Fig.~\ref{fig:Speckle}. 
If the source is pointlike, the intensity at the observation plane is uniform.
If the source has a finite size and consists of many parts emitting radiation with a definite wavelength $\lambda$ but with random relative phases, a ``speckle" pattern appears at the observation plane consisting of areas of higher an lower intensities as shown in the figure.
The typical size of a speckle spot is $d_{speckle}=\lambda {L}/{s}$ so that an observation of the speckle pattern yields an estimate of $s/L$, the angular size of the source.  
Compared to stellar interferometry, the HBT method is much less sensitive to fluctuations in the optical paths to the detector. 
\vskip 6pt
\noindent
{\it Aside}: An accurate estimate of speckle size is rather involved, (see  Ref.~\cite{goodman_speckle_2007} for a careful discussion), but a simple way to understand the characteristic size is to view a finite source as consisting of many pairs of point sources, each producing Young's fringes at the detection plane which must be summed.
The narrowest possible fringes correspond to two sources on opposite sides of the source, that is $\lambda L /s$. 
The intensity of the speckle pattern cannot spatially vary more rapidly than that. 
\vskip 6pt

As in the stellar interferometer, observing the speckle of a star is difficult because of its very short correlation time. 
The eye or any other slow detector will simply observe the average intensity everywhere at the detection plane.
Modern detector technology can achieve temporal resolutions in the 100 ps range, and therefore offers more hope. 
A 100~ps detector will attenuate 1~fs fluctuations of starlight by a factor of $10^{-5}$ and thus the intensity correlation between different points on the detection screen can be observed with enough averaging. 

In terms of electric fields and their correlations, one can describe the observation as follows: two detectors placed at positions $\mathbf r$ and $\mathbf{r+d}$ produce photocurrents $i(\mathbf{r},t_1)$ and $i(\mathbf{r+d}, t_2)$. 
The photocurrents are proportional to the light intensity $I$ received at the detector.
An electronic correlator produces a signal  $G^{(2)}(\mathbf d)\propto \langle i(\mathbf{r},t_1) i(\mathbf{r+d}, t_2)\rangle$.
In terms of the fields, the 2nd order correlation function is:  
\begin{equation}
G^{(2)}(\mathbf d, t_1, t_2)=\langle 
E^-(\mathbf{r},t_1)  E^+(\mathbf{r},t_1))   E^-(\mathbf{r+d},t_2 ) E^+(\mathbf{r+d},t_2)
\rangle.
\end{equation}
The stationary property of the source fields again implies that the correlation $G^{(2)}$ only depends on $\tau=t_1-t_2$, and henceforth we will only use one time variable $\tau$ in the correlation functions.
The measurement amounts to using the fact that for a speckle pattern, the light intensities at two points separated by a distance greater than $d_\mathrm{speckle}$ fluctuate in an uncorrelated way so that $\langle i(\mathbf{r},t) i(\mathbf{r+d}, t)\rangle=\langle i(\mathbf{r},t)\rangle \langle i(\mathbf{r+d}, t)\rangle$, while for negligible separations $\langle i(\mathbf r,t)^2\rangle \ge \langle i(\mathbf r,t)\rangle^2$.
The observation can also be carried out in counting mode, that is the arrival of individual photoelectrons are recorded and one counts coincident photoelectrons within a temporal window greater than or equal to the time resolution of the photodetector. 

The angular resolution achieved by the HBT method is thus entirely determined by the maximum separation $d$ over which one can observe correlations. 
Remarkably, this resolution is identical to that obtained by stellar interferometry.
The similarity stems from the two relevant correlation functions $G^{(1)}(\mathbf d,\tau)$ and $G^{(2)}(\mathbf d,\tau)$ and the nature of the light source used in the measurements. 
Thermal radiation is incoherent in the sense the complex electric field $E^+$ fluctuates  over all possible phases and over a range of amplitudes which determine the average intensity.
In the complex plane, the possible values of $E^+$ occupy a two-dimensional Gaussian distribution centered at zero. 

A powerful theorem concerning stochastic functions, known as Isserlis theorem \cite{munthe-kaas_short_2025} can be used to show that for a zero mean Gaussian field, the higher order correlation functions are simply related to the 1st order correlation function.
In particular when correlation functions are written in normalized form: $g^{(k)}((\mathbf d,\tau))=G^{(k)}(\mathbf d,\tau)/G^{(1)}(0,0)^k$, Isserlis theorem gives:
\begin{equation}  
g^{(2)}(\mathbf d, \tau)=1+ \abs{ g^{(1)}(\mathbf d, \tau)}^2.
\label{e:siegert}
\end{equation}
Thus we have $g^{(1)}(0,0)=1$ and therefore $g^{(2)}(0,0)=2$.
Even more important is the conclusion that the distance $d$ over which stellar interference fringes are lost is the same as that over which the \HBT correlation is lost. 
Application of Isserlis theorem to higher order correlations yields the interesting result that for zero spatial and temporal separations:
\begin{equation}
g^{(n)}=n! \,.
\label{e:Nfactorial}
\end{equation}

\section{Quantum theory}
The \HBT observation seemed difficult to understand in the context of an electromagnetic field described by photons. 
As mentioned in the introduction, if one imagines photons as ordinary particles, it is difficult to imagine how their detection can be correlated.
Fano \cite{fano_quantum_1961} and Glauber \cite{glauber_photon_1963} were among the first to emphasize that one must carefully model the process of photon detection in order to give a clear explanation.

A common model, which corresponds to most photodetectors, is to consider a bound electron (in an atom, a quantum well, a p-n junction {\it etc.}) which is excited to an unbound state by an oscillating electric field. 
We describe the interaction using the electric dipole approximation and therefore ignore the magnetic field in what follows. 
The unbound electron is then observed by some amplification process (such as a charge avalanche in a semiconductor).
Furthermore one must recognize that ``photons" are more accurately described as excitations of quantum fields and one must carefully treat the operator corresponding to the electric field. 
Assuming periodic boundary conditions in a large quantization volume $V$, the electric field operator can be expressed as:
\begin{equation}
\hat {E} =\frac{i}{2} \sum_{i}   \sqrt{\frac{2\hbar\omega_i}{\epsilon_0 V}}
(\hat a_i e^{i(\mathbf k_i \mathbf r - \omega_i t)}
-\hat{a}^{\dag}_i e^{-i(\mathbf k_i \mathbf r - \omega_i t)})
\label{e:field}
\end{equation}
where $\hat a^\dag_{k_i}$ and  $\hat a_{k_i}$ are creation and annihilation operators for plane waves with wave vector $\mathbf k _i$.
Here again we will neglect the vector nature of the field.
In analogy to the classical notation, we will denote $\hat{E}^+$ as the operator containing all the annihilation operators (the positive frequency part of the field) and $\hat{E}^-$ as the operator containing all the creation operators (the negative frequency part of the field). 

The probability of photo-detection by a single bound electron is governed by the matrix element of the electric field operator connecting an initial state, $i$ of the field to a final state $f$:
\begin{equation}
\bra{f} \hat {E}^+(\mathbf r,t) \ket{i}
\end{equation}
The ionization process must increase the electron energy and therefore must decrease the field energy. 
Therefore only the positive frequency part of the field, which contains photon annihilation operators, contributes. 
The ionization probability is proportional to the squared modulus of this amplitude summed over all final states.
A completeness relation allows one to eliminate the sum over final states and express the probability as an expectation value of the initial state of the system:
\begin{equation}
\mathrm{probability}\propto \bra i \hat {E}^-(\mathbf r,t) \hat {E}^+(\mathbf r,t) \ket i
\end{equation}
This relation is the obvious generalization of the quantity $G^{(1)}(0,0)$ in Eq.~\ref{e:classical_g1}. 
In an interferometry experiment, the same reasoning as in the classical case shows that the interference signal for a quantum field is given by 
\begin{equation}
G^{(1)}(\mathbf r_1 , \mathbf r_2,\tau) = \bra{i}\hat {E}^-(\mathbf r_1,t_1) \hat {E}^+(\mathbf r_2,t_2) \ket{i}
\end{equation}

In order to observe correlations, one must use two photo detectors, and the relevant amplitude is:
\begin{equation}
\bra{f}\hat {E}^+(\mathbf r_2,t_2) \hat {E}^+(\mathbf r_1,t_1) \ket{i}.
\end{equation}
Note the presence of two positive frequency operators, as required by energy conservation.
Once again this sum can be expressed as an expectation value of the initial state of the field, so that the probability is proportional to:
\begin{equation} 
G^{(2)}(\mathbf r_1,\mathbf r_2 ,\tau) =
\bra{i}
\hat {E}^-(\mathbf r_1,t_1) \hat {E}^-(\mathbf r_2,t_2)\hat {E}^+(\mathbf r_2,t_2) \hat {E}^+(\mathbf r_1,t_1)
 \ket{i}.
\end{equation}
Unlike in the classical case, the order of the operators here is important. 
For a classical field, the positive and negative frequency parts of the field contain the same information, while in the quantum case, one of them contains annihilation operators and the other creation operators.
Annihilation and creation operators for the same mode do not commute and this fact encapsulates most if not all the differences between the classical and quantum treatments of photon detection. 
In what follows, we will follow the classical discussion of Sec.~\ref{sec:classical}, denoting $\mathbf r_1$ as $\mathbf r$, $\mathbf r_2$ as $\mathbf r + \mathbf d$, $t_1$ as $t$, $\mathbf t_2$ as $t+\tau$, assuming translational invariance and stationarity of the correlations, and writing the correlations as functions of $\mathbf d$ and $\tau$ only.

It is often the case that the electric field is not in a pure quantum state, but rather a mixed state described by a density matrix $\rho$.
In this case the above correlation functions must be generalized to: 
\begin{equation}
G^{(1)}(\mathbf d,\tau) = 
Tr (\rho \, \hat {E}^-(\mathbf r,t) \hat {E}^+(\mathbf r+\mathbf d, t+\tau) )
\end{equation}
and
\begin{equation}
G^{(2)}(\mathbf d,\tau) = 
Tr (\rho \, \hat {E}^-(\mathbf r,t) \hat {E}^-(\mathbf r+\mathbf d, t+\tau) \hat {E}^+(\mathbf r +\mathbf d, t + \tau)\hat {E}^+(\mathbf r,t) ).
\end{equation}

It is instructive to examine the behavior of these correlation functions in a few simple cases.
We will describe the field in the basis of number states, that is the eigenstates $\ket {n_i}$ of the number operator for each mode, $\hat n_i \equiv \hat{a}^\dag _i \hat{a}_i$, with eigenvalue $n_i$.
Furthermore, we will concentrate on the normalized form of the correlations as in Eq.~\ref{e:siegert}.

First we consider thermal source such as a blackbody at temperature $T$. 
A mode of the field with frequency $\omega$ is described by a diagonal density matrix
\begin{equation}
\hat \rho = (1-e^{-\gamma})^{-1}
  \begin{pmatrix}
  1&0&0&...\\
  0&e^{-\gamma}&0& ...\\
  0&0&e^{-2 \gamma }& ...\\
  \vdots&\vdots&\vdots&
  \end{pmatrix}
\end{equation}
where we have defined $\gamma = \hbar \omega/k_B T$.
The mean photon number in that mode is $\langle \hat n \rangle = Tr(\hat \rho \, \hat n) = (e^\gamma - 1)^{-1}$, and one can easily show that $\langle\hat n^2 \rangle=2\langle \hat n \rangle^2$.
(In this section we use $\langle \rangle$ to denote the quantum mechanical expectation value.)
For a single mode, the spatial and temporal dependence disappears and we find:
\begin{equation}
G^{(1)} = \mathcal E^2 \langle \hat n \rangle ...  \; \mathrm{and}
\;  G^{(2)}  =
2  \mathcal E^4  \langle \hat n \rangle^2, 
\end{equation}
where we have defined $\mathcal E=\sqrt \frac{\hbar \omega}{2 \epsilon_0 V}$.
We thus confirm the classical results $g^{(1)}= 1$ and $g^{(2)}= 2$.
Indeed it is possible to prove an analog of Isserlis theorem for quantum operators.
Known as Wick's theorem (see Ref.~\cite{cohen-tannoudji_mecanique_2017} for a detailed discussion), this theorem allows one to show that for a single mode thermal state, the higher order correlation functions can be expressed in terms of the first order correlation function.
For a thermal state, the relation Eq.~\ref{e:Nfactorial} also holds.

A second simple example is that of a single mode coherent state.
A coherent state is the quantum state which most closely reproduces a monochromatic classical wave. 
The coherent state $\ket \alpha$ is an eigenstate of the annihilation operator $\hat a$ with complex eigenvalue $\alpha \equiv \abs{\alpha} e^{i \phi}$. 
From the definition of the field in Eq.~\ref{e:field}, we have 
\begin{equation}
\bra \alpha \hat E \ket \alpha =\mathcal E \abs{\alpha} \, \mathrm{sin}(\mathbf {k  r} + \omega t -\phi).
\end{equation}
We also find:
\begin{equation}
G^{(1)} =\mathcal E^2 \abs{\alpha}^2  \; \mathrm{and}
\;  G^{(2)}  =
\mathcal E^4 \abs{\alpha}^4
\end{equation}
Therefore, unlike a thermal state, for a single mode coherent state we have $g^{(2)}= 1$.
Wick's theorem does not apply in this case because, a coherent state with non-zero amplitude is not a zero-mean Gaussian state. 
The result $g^{(2)}= 1$ can also be easily generalized to a multifrequency coherent state.

One more situation of interest are the correlation functions of number states $\ket n$. 
Here again we do not expect Wick's theorem to apply, and indeed we find:
\begin{equation}
G^{(1)} = \mathcal E^2 n   \; \mathrm{and}
\;  G^{(2)}  =
\mathcal E^4  ( n ^2 -  n ).
\end{equation}
In the interesting case of $n=1$ we see than  $G^{(2)}=0$. 
This result is intuitive if one recalls that $G^{(2)}$ describes the correlation between two photon detections which has zero probability if there is only one photon in the state. 
This feature of the single photon state, which is the crux of the ``photon antibunching" effect, plays an important role in quantum optics as we shall see in the next section.

\section{Examples from experiments}

\subsection{Experiments with photons}

\begin{figure}[htbp]
\begin{center}
\includegraphics[width=12cm]{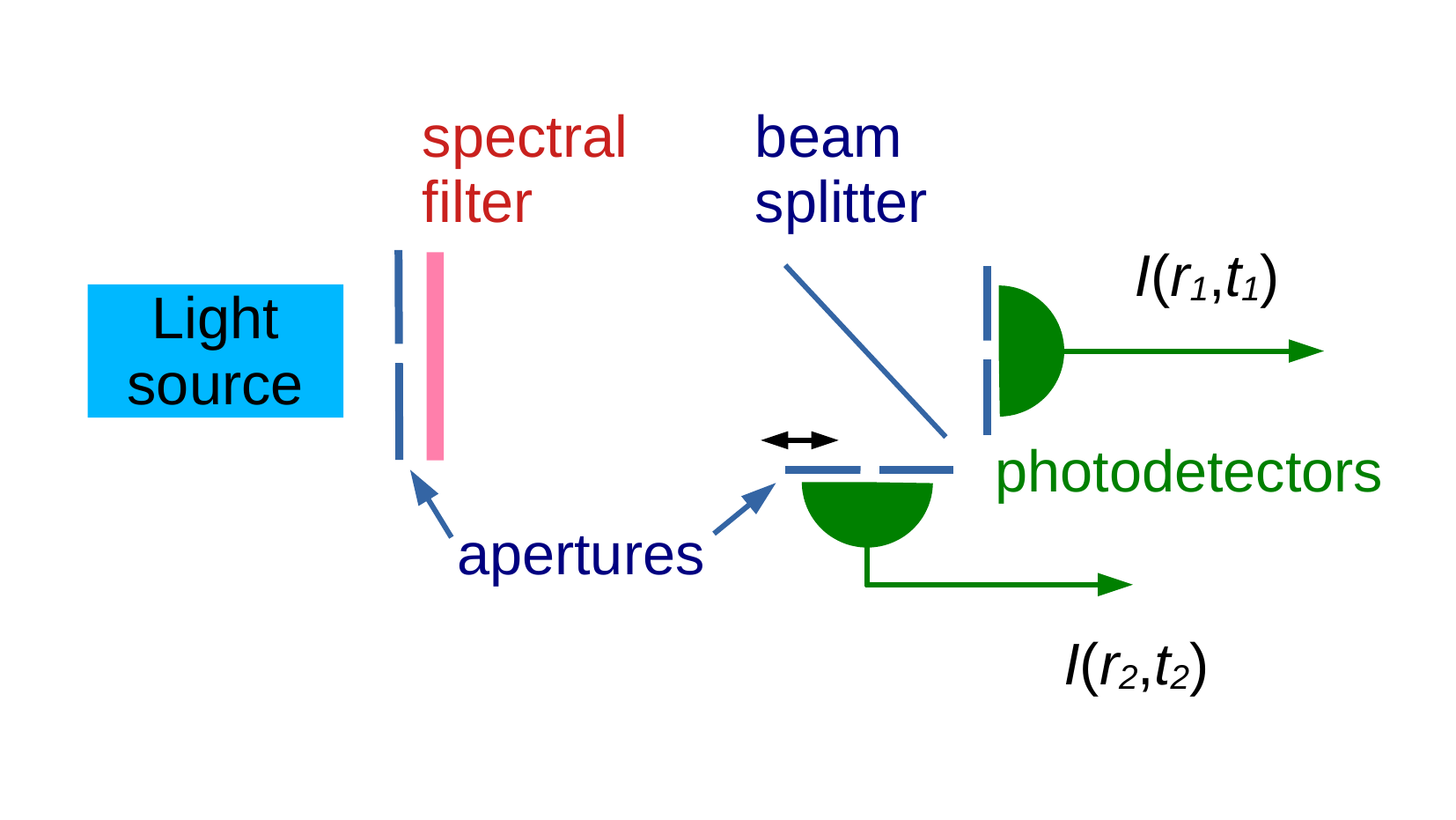}
\caption{Diagram of the Hanbury Brown and Twiss laboratory experiment.
The image of one detector is superposed with the other. The detector and its aperture are then moved to vary the spatial separation.}
\label{f:HBTExpt}
\end{center}
\end{figure}

In order to convince skeptics about the feasibility of their experimental proposal \HBT published the results of a series of laboratory experiments in which they measured correlations from light emitted by a spectral lamp \cite{hanbury_brown_intensity_1974,hanbury_brown_correlation_1956}. 
A diagram of the laboratory setup is shown in Fig.~\ref{f:HBTExpt}.
Unlike in the astrophysical experiments, the laboratory experiment used a beam splitter to send light to the two detectors.
The length scales in the laboratory experiment meant that the typical speckle size was smaller than the detector. 
Thus apertures were placed in front of each detector and the beam splitter superposed one detector with the image of the other one, permitting zero or very small separations $d$.
The use of 2 separate detectors also avoids the problem of distortion of the correlation due to detector deatime.  
Correlations were detected as a function of the transverse position of one of the detectors.

\begin{figure}[htbp]
\begin{center}
\includegraphics[width=12cm]{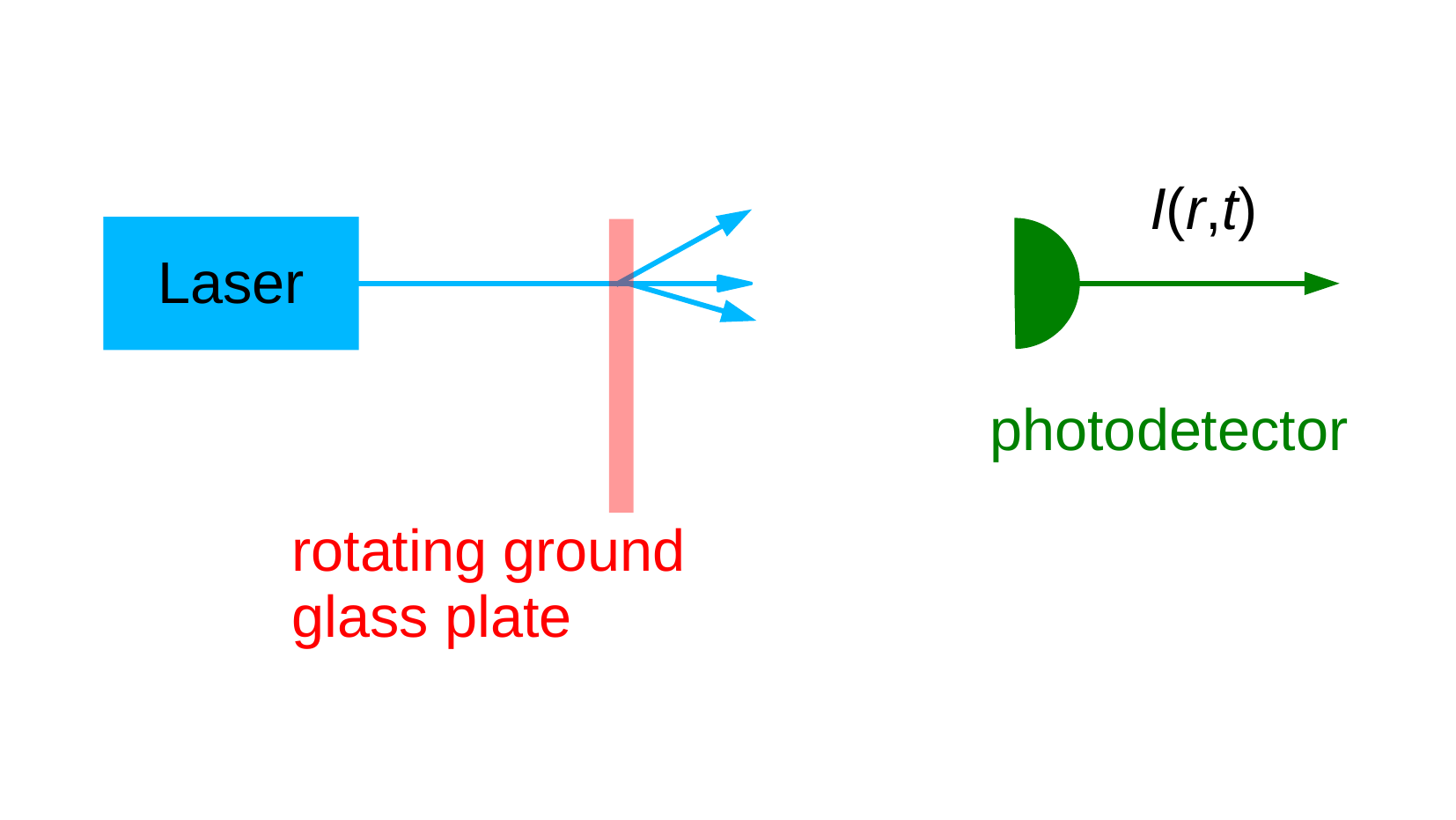}
\caption{Experiment to illustrate the difference between a laser and a thermal source. 
Two configurations are used.
In the first, a weak laser beam directly illuminates the detector. 
In the second, a rotating, ground glass screen is placed in the laser beam, scattering the light in a random fashion creating a speckle pattern. 
The motion of the screen causes the speckle pattern to fluctuate.
The detector then records the photon arrival times.
The rate for detecting pairs of photons as a function of their temporal separation is shown in Fig.~\ref{f:ArecchiData}.
}
\label{f:ArecchiExpt}
\end{center}
\end{figure}

\begin{figure}[htbp]
\begin{center}
\rotatebox[origin=c]{1}{
\includegraphics[width=8cm]{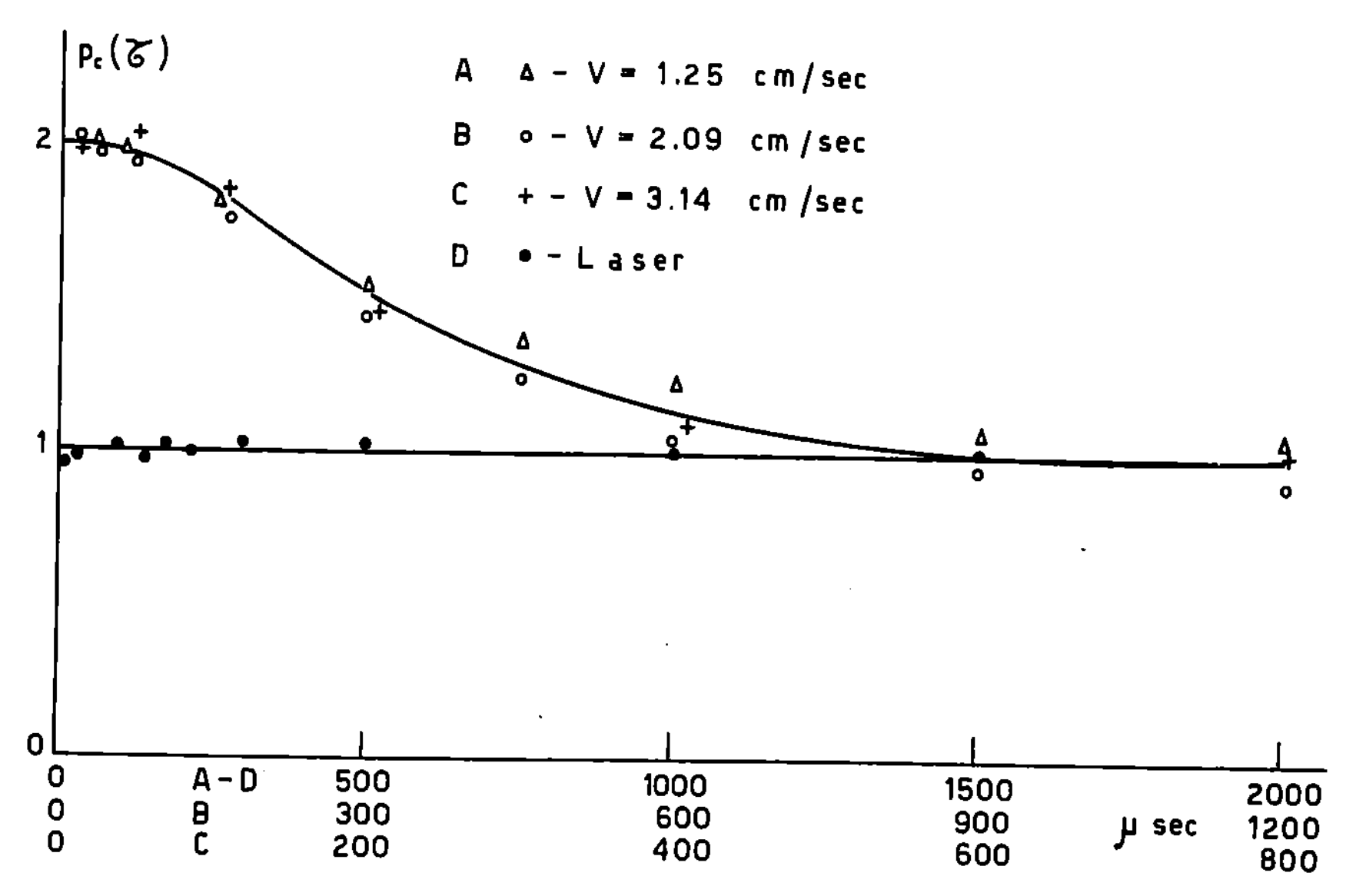}
}
\caption{Relative probability of detecting a second photon a time $\tau$ after having detected the first photon for a laser (flat line, curve D) and a speckle source (curves A, B, and C) using the apparatus sketched in Fig.~\ref{f:ArecchiExpt}. 
Figure is from Ref.~\cite{arecchi_time_1966}.
}
\label{f:ArecchiData}
\end{center}
\end{figure}

After the invention of the laser it was natural to investigate the photon correlations in such a light source. 
Figure \ref{f:ArecchiExpt} shows an experiment in which a single mode laser illuminates a photodetector \cite{arecchi_time_1966}. 
This experiment investigated photons detected at different times rather than at different positions and no second detector was necessary.
Detector deadtimes were unimportant because they were much shorter than the typical photon separations.
The experiment was conducted in photon counting mode and the correlation corresponded to the probability of detecting a second photon at a time $\tau$ after the detection of the first one.  The data is plotted in Fig.~\ref{f:ArecchiData}. 
One sees that the normalized correlation function is flat, that is, independent of the time delay between the detection of two photons.
On the other hand, a Gaussian field can be produced from the same laser by inserting a rotating, ground glass plate. 
The ground glass produces a speckle pattern at the detector and its rotation causes the speckle to fluctuate: the source is sometimes referred to as being ``quasi-thermal".
The time scale of the fluctuations is given roughly by the size of the laser beam spot divided by linear speed of the screen at the spot.
The data in Ref.~\ref{f:ArecchiData} shows that for a ground glass screen, the normalized correlation has the value of 2, decreasing to unity for large time separations, and that the correlation time is inversely proportional to the screen velocity.

The experiment of Ref.~\ref{f:ArecchiData} compared the light from a single mode laser with that in a fluctuating speckle pattern, that is a coherent state with a quasi-thermal one.
Both of these sources are ``classical" in the sense that they are easily described in the context of classical statistical optics.

\begin{figure}[htbp]
\begin{center}
\includegraphics[width=7cm]{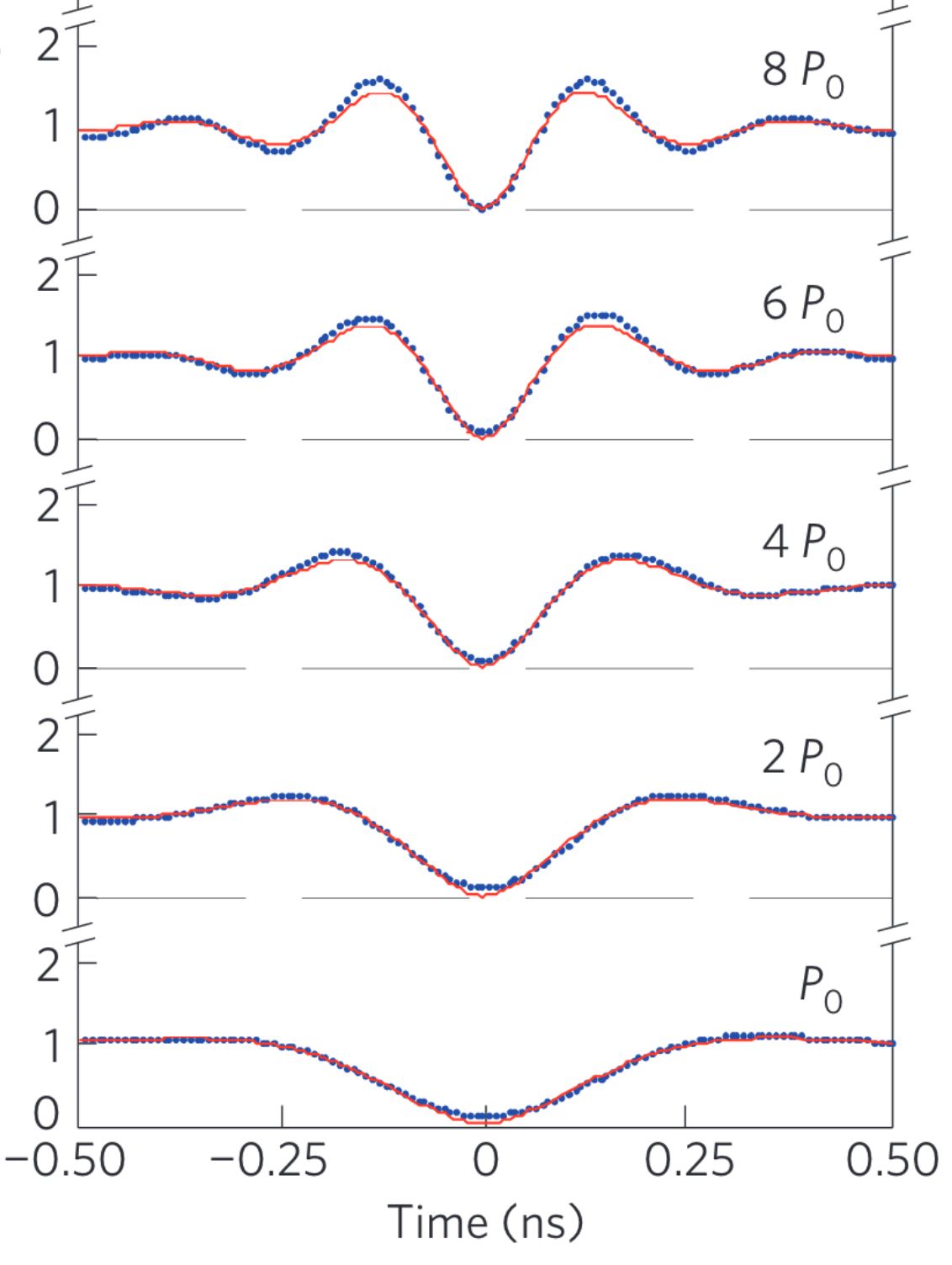}
\caption{Experimental measurement of the second-order correlation signal of light coming from a single quantum dot showing anti-bunching at coincidence (zero time delay). The traces show different pump laser powers, all of which are multiples of the lowest power $P_0$. The oscillations at larger delay correspond to Rabi oscillations between the two energy levels: after 1/2 an oscillation the atom is more likely than average to be in the excited state and emit a second photon thus enhancing the correlation.
As the pump power is increased these oscillations increase in frequency. 
The solid line is a fit to a model. 
Figure adapted from Ref.~\cite{flagg_resonantly_2009}.
}
\label{f:antibunching}
\end{center}
\end{figure}

About a decade after these experiments, L. Mandel and collaborators, perfected a non-classical source which realized a good approximation to the light emitted by a single atom excited by a continuous light beam. 
The source was a weak beam of sodium atoms in which, on average, fewer than one atom was inside the excitation volume at any given time.
The detector setup was very close to that used in the first HBT laboratory experiment, except that the correlation was observed as a function of the time difference.
Instead of observing a peak in the correlation at short times, the opposite effect, a dip was observed \cite{Kimble_photon_1977}.
Since that time, this type of measurement has been repeated many times to evaluate the quality of single photon sources.
Figure \ref{f:antibunching} shows a recent example.
The degree to which the correlation at zero delay deviates from zero gives an upper bound for the probability that the source emits more than one photon \cite{Lounis_single-photon_2005, Eisaman_invited_2011}. 


\subsection{Experiments with atoms}
The interference and correlation phenomena described above are not unique to the electromagnetic field. 
As early as 1960, and apparently independent of Hanbury Brown and Twiss, the effect was identified as contributing to observations of pions emitted in nuclear collisions \cite{goldhaber_influence_1960}.
The use of such effects in the nuclear physics community has been extensive and has yielded information about the length scales over which nuclear collisions take place \cite{baym_physics_1998, lisa_femtoscopy_2005}.

In the 1990's, progress in the laser cooling of atoms produced sources of atoms with coherence lengths and times long enough to be easily accessible in experiments. 
Interference experiments with atoms open new possibilities for precision measurements.
In particular, the atoms' mass renders such interferometers highly sensitive to inertial effects: rotation, acceleration and gravitation \cite{cronin_optics_2009}. 
The application of first order interference concepts to atoms and molecules is a vast field which we will not attempt to review here, but the availability of cold atoms also permitted experimenters to address fundamental questions such as differences between bosons and fermions, the coherence properties of Bose-Einstein condensates, and the effects of non-linearities induced by atom-atom interactions. 
In addition, cold atom manipulation and detection methods allow measurements in 3 dimensions, and of higher order correlations.

\begin{figure}[htbp]
\begin{center}
\includegraphics[width=0.4 \linewidth]{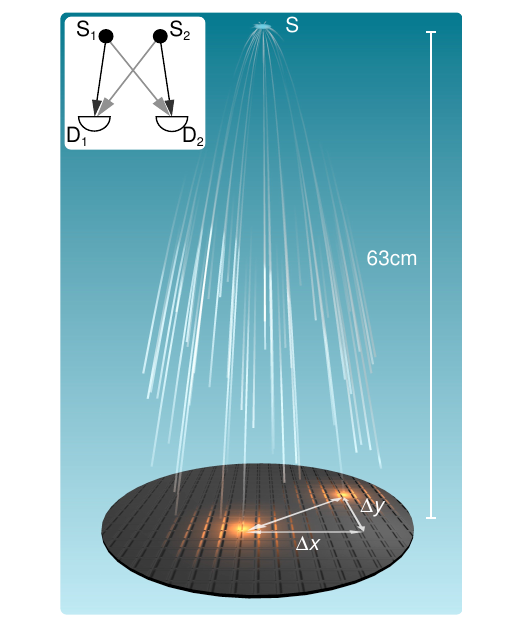}
\caption{Atomic HBT experiment.
Atoms from the source S fall onto an MCP (black disk).
The positions and arrival times are recorded and a histogram of pair separations $(\Delta x, \Delta y, \Delta z)$ is computed.
The inset shows possible paths between 2 source points and 2 detection points, illustrating  an argument discussed in Sec.~\ref{s:hom}.
Figure from Ref.~\cite{schellekens_hanbury_2005}.}
\label{f:AtomicHBTExpt}
\end{center}
\end{figure}

The atom experiments mentioned above function in close analogy with the operation of an astronomical observation: the atoms propagate a large enough distance to the detector so the speckle size is typically larger than the effective pixel size of the detector - no beam splitter is needed.
An interesting difference from the optical case is the effect of the acceleration of gravity (see Fig.~\ref{f:AtomicHBTExpt}).
This causes the deBroglie wavelength $h / p$ of the atoms to vary, so that this wavelength itself is not immediately related to the speckle size.
(The optical analog would be a continuously increasing index of refraction.)
A simple way to estimate the speckle size is to appeal to the uncertainty principle \cite{gomes_theory_2006}.
In the trap, the coherence length (or the thermal deBroglie wavelength) is related to the width of the momentum distribution $\Delta p$ by $\hbar/\Delta p$.
Analogously, the ``coherence momentum" $p_\mathrm{coh}$ is related to the size of the source $s$ by $p_\mathrm{coh}=\hbar/s$. 
After an expansion of duration $t$, the positions of the particles reflect their initial momenta and the speckle size at the detector is given by $d = (p_\mathrm{coh}/m) t =\hbar t / (m s)$.
For atoms suddenly released from a trap, this analysis is valid in all three spatial dimensions. 
If the trap is anisotropic, so too are the observed correlation lengths at the detector.
The situation is slightly different when one uses a continuous beam of atoms. 
In this case the transverse coherence lengths behave as described above, but the longitudinal coherence length is not determined by the source size but rather by the energy width of the source  \cite{gomes_theory_2006,yasuda_observation_1996}.
The continuous beam experiment of Ref.~\cite{yasuda_observation_1996} was the first to exhibit the atomic HBT effect and is even more closely analogous to the optical case in that the longitudinal correlation in the optical case is also determined by the energy width of the source. 

\begin{figure}[htbp]
\begin{center}
\includegraphics[width=0.7 \linewidth]{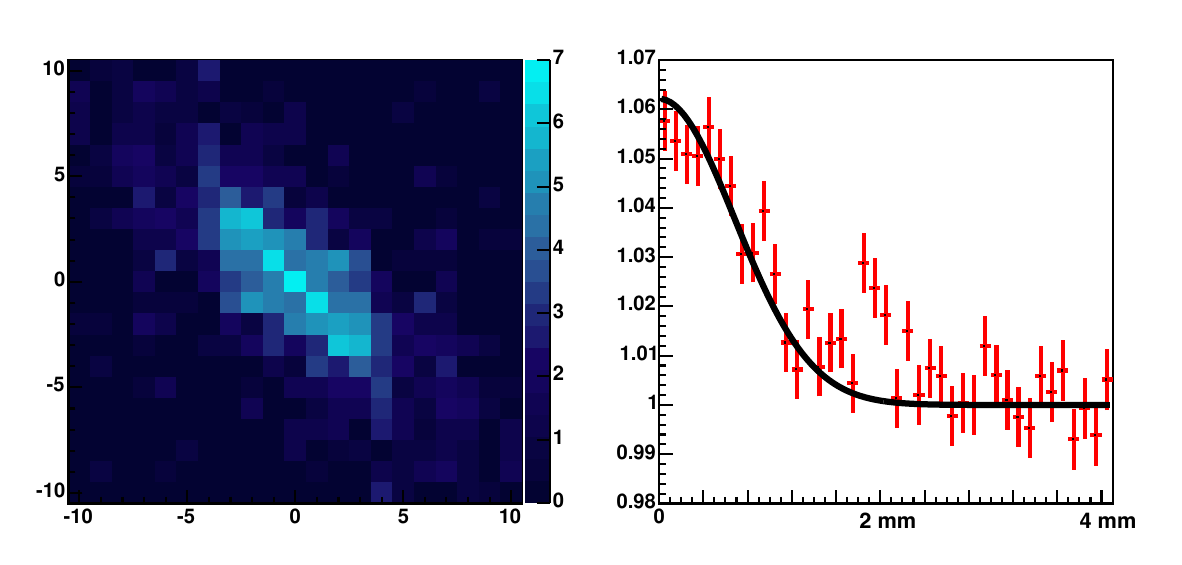}
\caption{Data showing the atomic HBT effect in 3D. 
Left: The observed, normalized pair correlation $g^{(2)}$ in the plane of the detector. 
The image is a histogram of pair separations of up to $\pm 2$ mm along both directions. 
The false color shows the value of $100 \times (g^{(2)}-1)$
Right: The histogram in the vertical direction. 
The solid line is a fitted Gaussian function, whose width $d \approx 0.8\,\mathrm{mm}$ is consistent with the known source size $s\approx 6\, \mu\mathrm{m}$.
The source temperature was about 1~$\mu$K.}
\label{f:AtomicHBTData}
\end{center}
\end{figure}

In Fig.~\ref{f:AtomicHBTData} we show the result of an experiment in which metastable helium atoms (in the $2^3 \mathrm S_1$ state) were released from a trap onto a microchannel plate (MCP) detector  \cite{schellekens_hanbury_2005}.
The figure shows the correlation both in the plane of the detector and in the direction orthogonal to it. 
The 2 dimensional plot shows the anisotropy due to the anisotropic trap shape.
The aspect ratio of the trap was over 20, in the correlation the aspect ratio is a factor of about 10 smaller because the transverse resolution was not sufficient. 
The lack of resolution along one axis also explains why the correlation only goes up to 1.06 instead of 2. 
The atomic HBT effect was also demonstrated in the case of a gas trapped in an optical lattice \cite{folling_spatial_2005}.
In this case a periodic pattern was observed whose period corresponded to the reciprocal lattice vector. 

\begin{figure}[htbp]
\begin{center}
\includegraphics[width=0.9\linewidth]{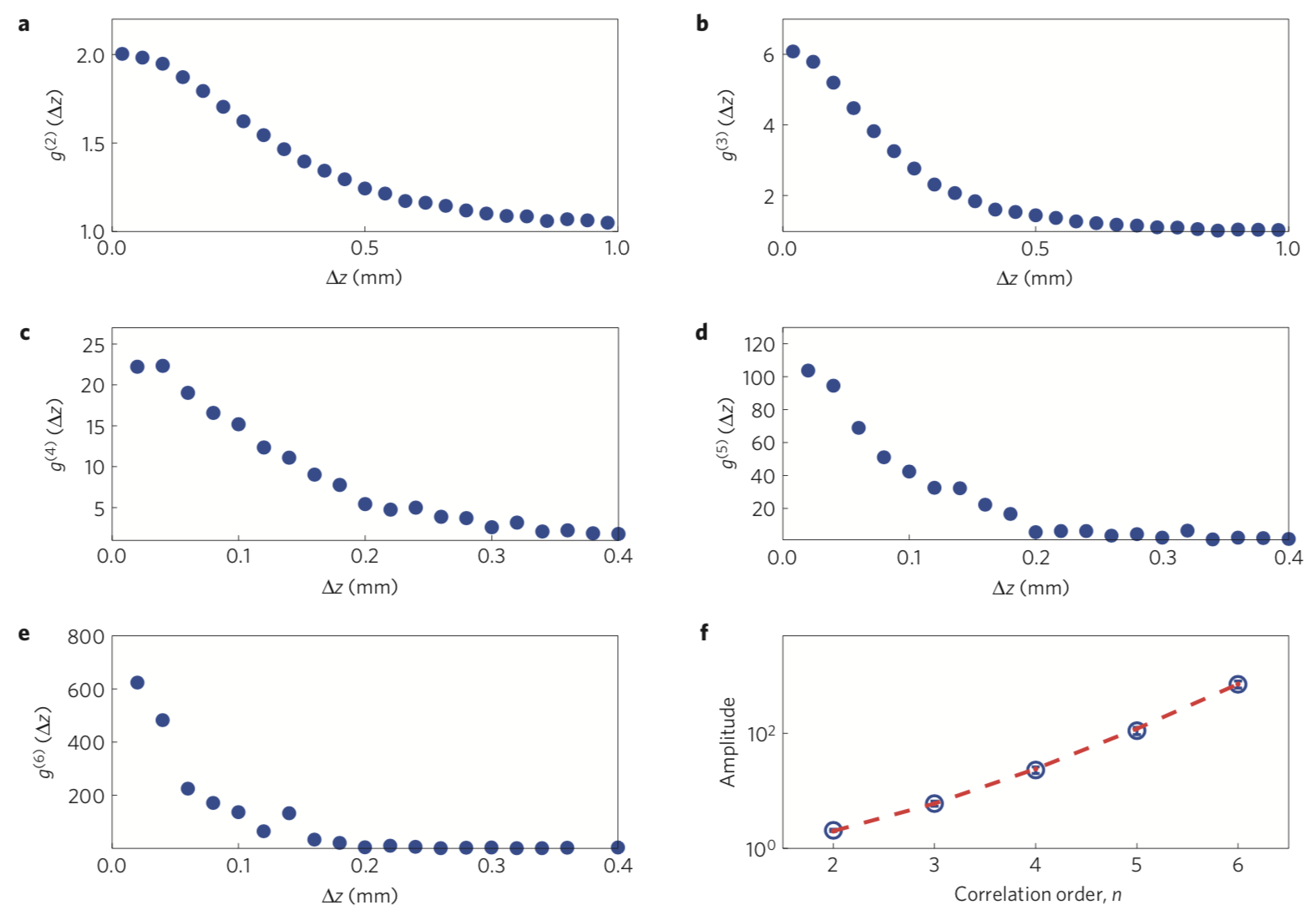}
\caption{Higher order correlation functions of a thermal Bose gas. 
The plots show increasingly high order correlation functions in a setup similar to Fig~\ref{f:AtomicHBTExpt}.
The plot at the lower right shows the amplitude of the correlation at zero separation as a function of its order.
Figure from Ref.~\cite{dall_ideal_2013}.
}
\label{f:Canberra}
\end{center}
\end{figure}

The excellent resolution and low background of the MCP detector has allowed experimenters to observe higher order correlations in a thermal Bose gas. 
Since one simply records the three coordinates of each atom, 3rd, 4th etc. correlations are straight forward, though time consuming, to compute. 
The example shown in Fig.~\ref{f:Canberra} shows correlations up to the 6th order. 
Unlike in Ref.~\ref{f:AtomicHBTData}, the initial size of the boson cloud and the detector resolution were well matched so that the speckle size was larger than the detector resolution in all three dimensions. 
The experiment demonstrates, as predicted by Wick's theorem, that the $n$th order correlation at zero distance is equal to $n!$.


\begin{figure}[htbp]
\begin{center}
\includegraphics[width=0.6\linewidth]{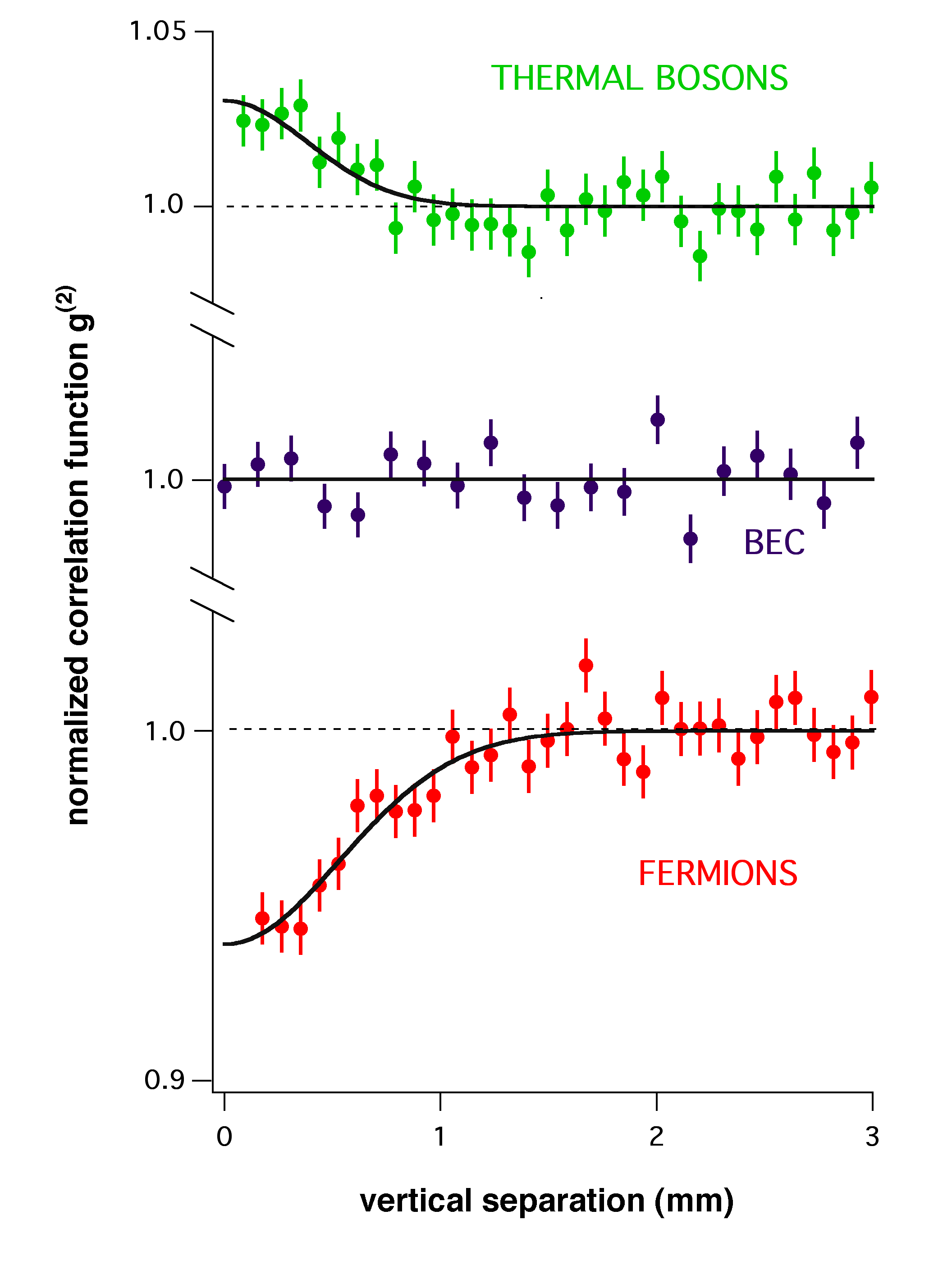}
\caption{Comparison of the second order correlation functions for thermal Bose and Fermi gases and a Bose-Einstein condensate.  
The vertical scale is the same in each graph.
Figure from Ref.~\cite{vassen_cold_2012}.
}
\label{f:fermibosebec}
\end{center}
\end{figure}

At the same time that methods to observe the HBT effect in a thermal Bose gas were being perfected, workers also considered going beyond this basic situation. 
Doing correlation experiments with atoms opens the possibility of using fermions instead. 
In this case one expects the exclusion principle to inhibit correlations at small distances.
The metastable helium atom is well adapted to such a project since one can use nearly the same apparatus to work with either $^4\mathrm{He}$ or $^3\mathrm{He}$, and indeed to perform the experiment simultaneously in the same vacuum chamber and with the same detector. 
Figure \ref{f:fermibosebec} shows a comparison of the correlation function along one direction for the two isotopes. 
Instead of bunching, one observes anti-bunching of fermions and over the same spatial scale $\hbar  t /(m s)$ as the bunching in the Bose case. 
The anti-bunching curve is slightly wider and deeper because of the smaller mass of the fermionic isotope. 
As in Fig.~\ref{f:AtomicHBTData}, neither does the bunching go to 2 nor the anti-bunching to 0 as in the ideal case because of the inadequate resolution of the detector along one of the source axes \cite{jeltes_comparison_2007}. 
Anti-bunching of atomic fermions has also been observed using absorption imaging techniques \cite{rom_free_2006}.

It is also possible to perform the experiment using a Bose-Einstein condensate (BEC). 
It is well known that a BEC shares some characteristics of the laser, especially in the sense that a large number of particles occupies a single mode of the system. 
Thus although unlike a laser, a BEC is an equilibrium state, both have similar correlation properties. 
This is shown clearly in Fig.~\ref{f:fermibosebec}  where it is seen that under conditions almost identical to the other data, the BEC exhibits a flat correlation function \cite{schellekens_hanbury_2005}. 
Later experiments studied the 2nd order correlation function of a Bose gas is it crossed the condensation threshold \cite{perrin_hanbury_2012}.

\newpage
\section{Hong Ou Mandel effect}
\label{s:hom}

\begin{figure}[htbp]
\begin{center}
\includegraphics[width=0.9\linewidth]{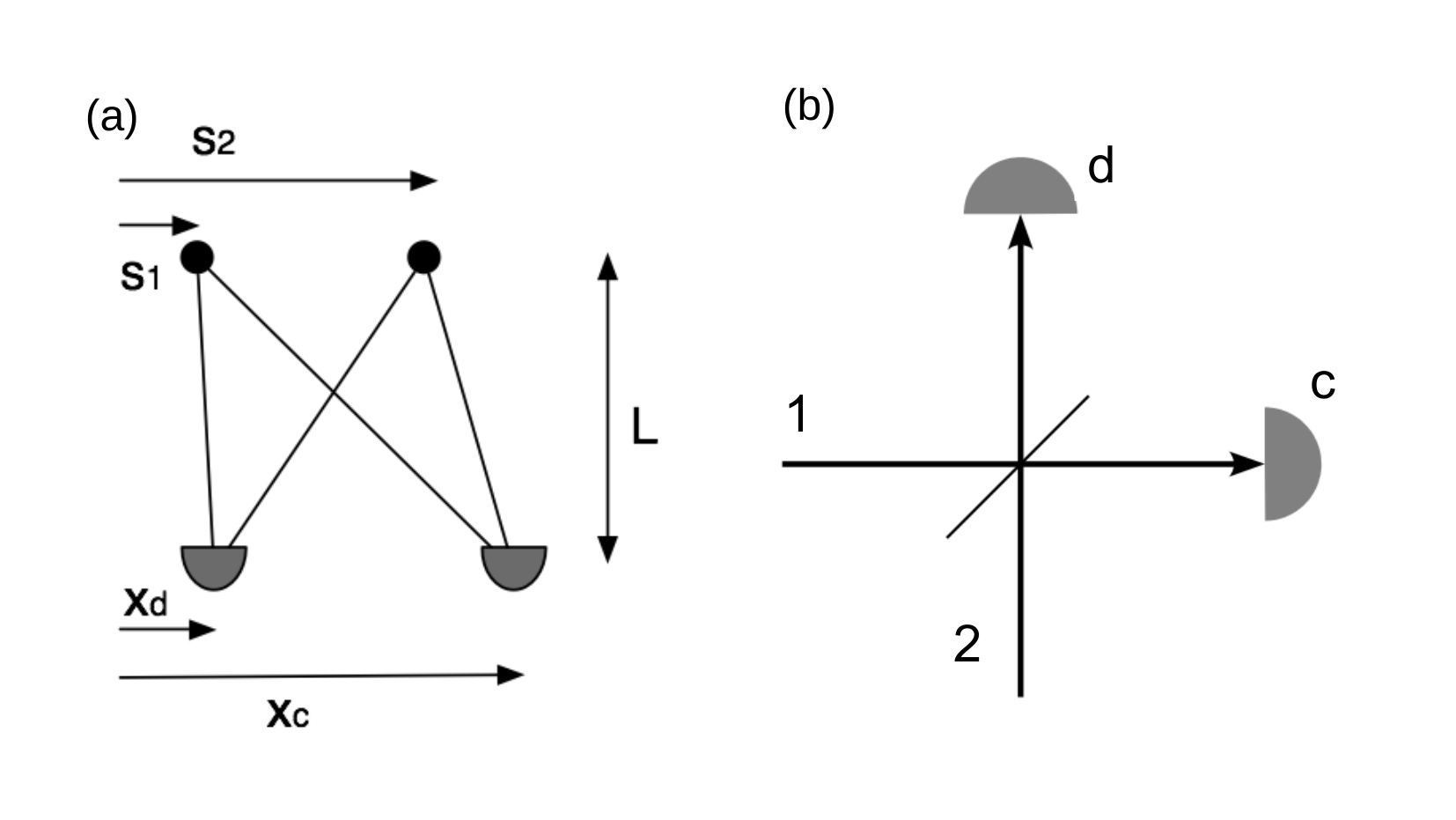}
\caption{Two configurations for observing 2 particle interference effects. 
(a) Two sources at $s_1$ and $s_2$ emit fields detected by detectors at $x_c$ and $x_d$. 
The joint detection probability at the two detectors exhibits Young's fringes even if the two sources are mutually incoherent. 
The sum over many source pairs yields the HBT signal. 
The Ghosh and Mandel experiment used  this configuration and a source of single photon pairs \cite{ghosh_observation_1987}.
(b) The configuration of the Hong Ou Mandel experiment \cite{hong_measurement_1987}.
A beam splitter was used to combine photons from sources 1 and 2 before detection at c and d. 
}
\label{f:GM-HOM}
\end{center}
\end{figure}

The Hong Ou Mandel (HOM) effect is a more dramatic two particle interference effect which in some cases does not have a classical explanation \cite{ou_multi_2007}. 
Like the \HBT effect it was first demonstrated for photons but was subsequently observed with other particles
We will show below that the step from HBT to HOM is a small one by appealing to an early discussion of the \HBT effect given by Fano \cite{fano_quantum_1961}.
The HBT effect involves the detection of two photons emitted by a source of finite size.
The simplest model of a finite sized source consists of two separated points, 1 and 2 which emit photons detected by detectors $c$ and $d$, see Fig.~\ref{f:GM-HOM}(a).
To detect one photon in detector $c$ and one in $d$, one of four processes must take place: i) source 1 emits two photons one of which is detected in $c$ and the other one in $d$, ii) source 2 emits two photons which are detected one in $c$ and one in $d$, iii) source 1 emits a photon detected in $c$ and source 2 emits a photon detected in $d$, iv) source 1 emits a photon detected in $d$ and source 2 emits a photon detected in $c$. 
Processes i) and ii) are distinguishable because it is in principle possible to tell which source emitted the photons. 
Therefore they do not produce any interference, but iii) and iv) constitute two indistinguishable ``paths" leading from the same initial state to the same final state. Thus, the associated quantum mechanical amplitudes can interfere if the two photons are indistinguishable.
If we model an extended source like a star as the sum over many source points the interference fringes wash out unless the detectors are sufficiently close together. 
This argument clearly illustrates how the quantum picture of the HBT effect relies crucially on the interference of two particle amplitudes, and we will develop an explicit quantitative model in Sec.~\ref{s:recovering}

\subsection{The Ghosh and Mandel effect}
\label{s:GMeffect}
Now we turn to the HOM effect, or rather to a closely related experiment performed by Ghosh and Mandel \cite{ghosh_observation_1987} in the same year as the  HOM experiment.
Consider again the sources and detectors described in Fig.~\ref{f:GM-HOM}(a) but instead of averaging over many sets of pairs, we think of exactly two, spatially separated, pointlike sources.
If the emitted fields are monochromatic coherent states, one would observe Young's fringes in the detector plane, that is each detector would separately record an oscillating detection probability as it is moved left and right.
On the other hand, the sources might be mutually incoherent.
One might consider for example two coherent fields with a random fluctuating relative phase.
In this case, one expects no Young's fringes, that is, no first order coherence.
The correlation between the two detectors however does exhibit an oscillation as a function of their separation. 

To see this, let us imagine a situation as in the diagram.
We suppose that the emitted fields correspond to operators $\hat E^+ _1$ and $\hat E^+ _2 \exp i \theta$, where $\theta$ is random on the interval $[0, 2\pi]$.
Taking into account the phase acquired during propagation to the detector, the field at detector $c$ is $\hat E^+ _c = \hat E^+ _1 \exp(i \phi_{1,c}) + \hat E^+ _2 \exp(i \phi_{2,c}+i \theta)$.
Here $\phi_{l,m} = k r_{l,m}$, $k=2\pi/\lambda$ and $r_{l,m}$ is the distance from source $l$ to detector $m$.  
To simplify the notation we ignore the $1/r$ attenuation of the fields as they propagate to the detectors.
The second order correlation function $G^{(2)}_{c,d}$ at the detection plane is given by the (quantum) expectation value
\begin{equation}
G^{(2)}_{c,d} =  \langle \hat E^- _c \hat E^- _d \hat E^+ _d \hat E^+ _c \rangle.
\label{e:G2}
\end{equation}
This function depends on the positions $x_c$ and $x_d$ of the detectors but we suppress these arguments here.
Written in terms of $\hat E^+ _1$ and $\hat E^+ _2$ the function has 16 terms, but after averaging over the random variable $\theta$, only 6 terms survive. 
Let $\langle \hat E^- _1 E^+ _1 \rangle = |\alpha|^2$ and $\langle \hat E^- _2 E^+ _2 \rangle = |\beta|^2$  be the squared amplitudes of the two source fields, we then find:
\begin{equation}
G^{(2)}_{c,d} = |\alpha |^4+|\beta |^4 + 2 |\alpha |^2|\beta |^2 (1+ \cos [\phi_{1,c}-\phi_{2,c}-\phi_{1,d}+\phi_{2,d}]).
\label{e:ClassicalHOM}
\end{equation}
The phases are given by:
\begin{align*}
\phi_{l,m}&=k r_{l,m}=k \sqrt{L^2+(s_l-x_m)^2}\\
&= k L(1 +\frac{1}{2}(s_l-x_m)^2 /L^2 
+ ...)
\end{align*}
We will assume that the higher order terms in the above expansion contribute negligibly to the phase, in which case the interference term can be written $\cos\left[ k (s_1 - s_2)(x_d - x_c)/L \right ]$. 
That is, the second order correlation function exhibits the same oscillation as ordinary Young's fringes except that the argument is the \emph{separation} of the two detectors rather than their absolute positions. 

We can also consider two source points each of which emits exactly one photon. 
In this case, regardless of the phase $\theta$, only 4 terms contribute to the correlation function and one finds:
\begin{equation}
G^{(2)}_{c,d} \propto 1+\cos\left[ k (s_1 - s_2)(x_d - x_c)/L \right ]
\end{equation}
Here again we find two-particle interference fringes \emph{but with 100\% contrast}.
This means that given a detection at position $x_c$, there are points at which the probability of detecting a second photon in detector $d$ vanishes. 
This effect was discussed and demonstrated in Ref.~\cite{ghosh_observation_1987}.
The arguments given above for the classical and the two-photon cases are essentially those of that paper. 
One way to interpret the 100\% contrast in the two photon case is to notice that the first two terms of \eqref{e:ClassicalHOM} correspond to one of the sources emitting more than one photon, something which is assumed to be impossible in the two photon case. 
If one repeats this calculation for sources emitting single fermions, one finds a coherence function 
\begin{equation}
\label{e:GM-Fermi}
G^{(2)}_{c,d} \propto 1-\cos\left[ k (s_1 - s_2)(x_d - x_c)/L \right ],
\end{equation}
that is, one sees the same fringes, but shifted by one half a period. 

\subsection{Recovering the HBT effect}
\label{s:recovering}

It is instructive here to consider again the relation of this effect to the \HBT effect. 
We will refine the qualitative argument made in the first paragraph of Sec.~\ref{s:hom} for the case of $N$ source points $s_i$ arranged in a line which is a distance $L$ from the two detectors. 
To simplify the calculation we assume that source points are monochromatic and emit light at the same frequency and the same mean intensity: the temporal aspect therefore disappears.
The correlation between the detectors at $x_c$ and $x_d$ is again given by Eq.~\ref{e:G2}.
The field at each detector is given by a sum over the source points, but now there are several:
\begin{equation}
\hat E^+_c = \mathcal E \sum_{j=1}^N \hat a_j e^{i \phi_{j,c}}
\end{equation}
The correlation is given by:
\begin{equation}
 \langle \hat E^- _c \hat E^- _d \hat E^+ _d \hat E^+ _c \rangle =
\mathcal E^4 \sum_{j,k,l,m} \exp[i(- \phi_{c,j} -\phi_{d,k}+\phi_{d,l}+\phi_{c,m})]
\langle \hat a^\dag _j \hat a^\dag _k \hat a _l \hat a _m  \rangle
\end{equation}

Let us now assume that the sources emit single photon states. 
In this case only terms which have the same number of creation and annihilation operators for each mode contribute and we have:
\begin{equation}
\langle \hat a^\dag _j \hat a^\dag _k \hat a _l \hat a _m \rangle =
\langle \hat a^\dag _j \hat a^\dag _k \hat a _j \hat a _k  \rangle  \delta_{j,l}\delta_{k,m} +
\langle \hat a^\dag _j \hat a^\dag _k \hat a _ k \hat a _j  \rangle  \delta_{j,m}\delta_{k,l} -
\langle \hat a^\dag _j \hat a^\dag _j \hat a _j \hat a _j  \rangle  \delta_{j,k}\delta_{j,l}\delta_{j,m}.
\end{equation}
Now we can use boson commutation relations and define $\hat n_j = \hat a_j ^\dag \hat a_j$ to find:
 \begin{align*}
G^{(2)}_{c,d} =& \mathcal E^4
\sum_{j,k} (\langle \hat n_j \hat n_k \rangle - \delta_{j,k}\langle \hat n_j \rangle)(1+\exp[i(- \phi_{c,j} -\phi_{d,k}+\phi_{d,j}+\phi_{c,k})] )\\ 
&-\delta_{j,k} (\langle \hat n_j^2 \rangle - \langle \hat n_j \rangle) ].
\end{align*}
We can simplify the argument in the exponential as in Eq.~\ref{e:ClassicalHOM}. 
If we now consider the limit of a large number of source points, the terms proportional to $\delta_{j,k}$ can be neglected and we find:
\begin{equation}
G^{(2)}_{c,d}= \mathcal E^4 \sum_{j,k} \langle \hat n_j \rangle \langle \hat n_k \rangle 
\left(1+\exp[i \frac{k}{L}(s_j - s_k)(x_c - x_d)]\right)
\end{equation}
If we now suppose that the source points can be described as a distribution $n(s)$, where $s$ is a position on the line containing the points, we can make the replacements
\begin{align}
\sum_j \langle \hat n_j \rangle &\rightarrow N\\
\sum_j \langle \hat n_j \rangle e^{i \frac{k}{L} s_j (x_c - x_d))} &\rightarrow 
N \int  n(s) e^{i \frac{k}{L} s (x_c - x_d) } d s = N \tilde n(x_c - x_d)
\label{e:largeN}
\end{align}
where $\tilde n$ denotes a Fourier transform. 
Thus we find 
\begin{equation}
G^{(2)}_{c,d}=\mathcal E^4 N^2(1 + \abs{\tilde n(x_c - x_d)}^2).
\label{e:HBT-HOM}
\end{equation}
We see that the 2nd order correlation function is a constant plus the squared modulus of the Fourier transform of the source distribution, and that the quantum nature of the single photon sources has been submerged by the  sum over the $N$ sources.

We can recover the relationship of of Eq.~\ref{e:siegert} by also computing the first order correlation function between the points $x_c$ and $x_d$:
\begin{equation}
G^{(1)}_{c,d} = \langle \hat E^- _c \hat E^+ _d \rangle = \mathcal E^2 
 \sum_{j,k} \exp[i(- \phi_{c,j} +\phi_{d,k})]
\langle \hat a^\dag _j \hat a _k  \rangle.
\end{equation}
Again, the assumption of single photon states means than $\langle \hat a^\dag _j \hat a _k  \rangle =\langle \hat n_j \rangle \delta_{j,k}$. 
The phases can be written as: 
\begin{equation}
- \phi_{c,j} +\phi_{d,j} = \frac{k}{2 L}\left (-x_c^2+x_d^2 + 2 s_j(x_c - x_d) \right),
\end{equation}
and we find:
\begin{equation}
G^{(1)}_{c,d}=\mathcal E^2 e^{i \frac{k}{2L}(-x_c^2+x_d^2)}\sum_j \langle \hat n_j \rangle 
e^{i \frac{k}{L}(s_j(x_c - x_d))}.
\end{equation}
In the limit of large $N$ we again make the same replacement as in Eq.~\ref{e:largeN}, which shows that the first order correlation function is proportional to the Fourier transform of the source distribution. 
Thus we find the normalized form:
\begin{equation}
g^{(1)}_{c,d} = {G^{(1)}_{c,d}}/ (\mathcal E^2 N) = e^{i\frac{k}{2L}(-x_c^2+x_d^2)} \tilde n(x_c -x_d)
\end{equation}
The normalized form for the 2nd order correlation is:
\begin{equation}
g^{(2)}_{c,d} =\frac{ \langle \hat E^-_c \hat E^-_d \hat E^+_d \hat E^+_c \rangle}
{ \langle \hat E^-_c \hat E^+_c \rangle \langle\hat E^-_d  \hat E^+_d \rangle}
= 1+ \abs{\tilde n (x_c - x_d) }^2
\end{equation}
and we find the same result as Eq.~\ref{e:siegert}.

%


\subsection{Using a beam splitter}
\label{s:HOMeffect}
Instead of a geometry akin to a Young's double slit experiment, one can consider a configuration in which a beam splitter, rather than spatial propagation, is used to superpose the fields from the two sources (Fig.~\ref{f:GM-HOM}(b)).
This geometry resembles the HBT laboratory geometry in that a beam splitter is used.
Here however, a finite sized source is replaced by two single mode sources and these sources are at the two different input ports of the beam splitter.
The situation is thus reduced to two input modes and two output modes.
This was the configuration used by HOM in a paper which followed the Ghosh and Mandel paper by only a few months \cite{hong_measurement_1987}. 
When each source emits a single photon, the effect is even more dramatic because of the complete absence of events in which one photon is detected in each detector. 
The two photons always end up both in detector $c$ or both in $d$.
The analysis proceeds in much the same way as in the Ghosh and Mandel experiment except that it is not necessary to keep track of spatially dependent phases such as $\phi_{1,c}$. 
A crucial point in both experiments is that the arrival time of the particles at the detectors be such that it is impossible to tell the origin of either photon. 
One hundred fs shifts in the arrival times resulted in a loss of the HOM interference signal.
When a classical source, such as two coherent pulses with no fixed phase relationship, is used, the cross correlation decreases but the contrast gives at best a 50\% contrast rather than 100\% as in the two photon case.

One interesting detail in the original Ghosh and Mandel and HOM experiments is the fact that they used photon pairs created in parametric downconversion. 
Energy conservation meant that the two photons were correlated in frequency and not strictly independent. 
In subsequent years the experiment has been repeated using truly independent sources of photons \cite{santori_indistinguishable_2002, beugnon_quantum_2006} so that no residual correlation exists except that which is imposed by quantum mechanical symmetry. 
The HOM effect has become important in the field of quantum information when it is necessary to ensure that a single photon source consistently emits photons in the same spatio-temporal mode. 
The HOM contrast allows workers to characterize the repeatibility of their sources \cite{Bouchard_two-photon_2021,Lal_indistinguishable_2022}. 

Another interesting point concerning the HOM effect is that, like photon bunching, the result does not follow simply from the  exchange symmetry of identical photons. 
It is possible to change the relative phase of the paths taken by the two photon state, and convert the HOM ``dip" into a ``peak" \cite{Kwiat_observation_1992, Vest_anti-coalescence_2017}
Thus photons do not necessarily exhibit a dip in the cross correlation function. 
Fermions on the other hand are required by their anticommutation relations never to occupy the same mode, thus an HOM experiment with fermions can only observe a peak. 

\begin{figure}[htbp]
\begin{center}
\includegraphics[width=0.7\linewidth]{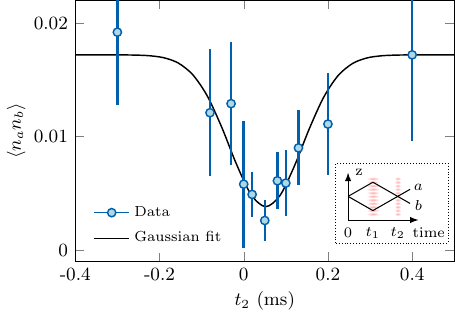}
\caption{Hong Ou Mandel experiment using metastable helium atoms.  
The inset shows the experimental configuration featuring one Bragg grating to deflect the atom trajectories at time $t_1$ and another which acts as a beam splitter at time $t_2$.
The correlation between the number of detected atoms in the two outputs $a$ and $b$ is shown as a function of the time of application of the beamsplitter.
The correlation is expressed as the product of the number of atoms detected in the two regions, averaged over many realizations.
Figure from Ref.~\cite{Perrier_thermal_2019}.
}
\label{f:atomic_hom_data}
\end{center}
\end{figure}

The atomic analog of the HOM experiment has also been performed. 
A close analog was performed using the same metastable helium apparatus as in the HBT experiments; in particular the atom detector was the same.
As in the original HOM experiment, a source which creates correlated atom pairs, in this case via a nonlinear atomic interaction in a BEC \cite{Bonneau_tunable_2013}.
After their creation, the two atoms separate along the vertical direction.
These atoms interact with a laser standing wave which acts as a Bragg diffraction grating, reversing their momenta so that they meet again at a later time. 
When they are superposed again, a second Bragg grating with a 50\% reflection probability, acts as a beam splitter with two outputs labeled $a$ and $b$ (see the inset of Fig.~\ref{f:atomic_hom_data}).
The atoms at the outputs of the beam splitter are detected and the correlation between the atom number in two outputs $\langle n_a n_b \rangle$ is calculated \cite{Lopes_atomic_2015}.
(Here the brackets denote an average over many repetitions of the experiment.)
The figure shows measurements of this correlation as a function of the time at which the beam splitter was applied. 
Interference, that is a reduction in the correlation, only takes place when the timing of the beam splitter is such that the paths taken by the atoms cannot be distinguished.
The width of the dip is related to the energy width of the source of atom pairs.   

The recent literature contains other examples of the HOM and the Ghosh and Mandel effect using atoms. 
Notable examples used pairs of atoms trapped in two optical optical potential wells to observe two particle interference effects. 
In \cite{Kaufman_two-particle_2014, Kaufman_hongoumandel_2018} the potential wells were sufficiently close that tunneling between them was possible, and thus the tunnel barrier acted like a beam splitter. 
The experiment showed that after one half of a tunnelling oscillation the two (bosonic) atoms had a tendency to be found both in the same well, rather than in different wells.
In other words, given two atoms initially on opposite sides of a tunnel barrier, and a duration corresponding to a 50\% probability of tunneling for each atom, the final state consists of a superposition of two atoms in either of the two wells.
 
\begin{figure}[htbp]
\begin{center}
\includegraphics[width=0.8\linewidth]{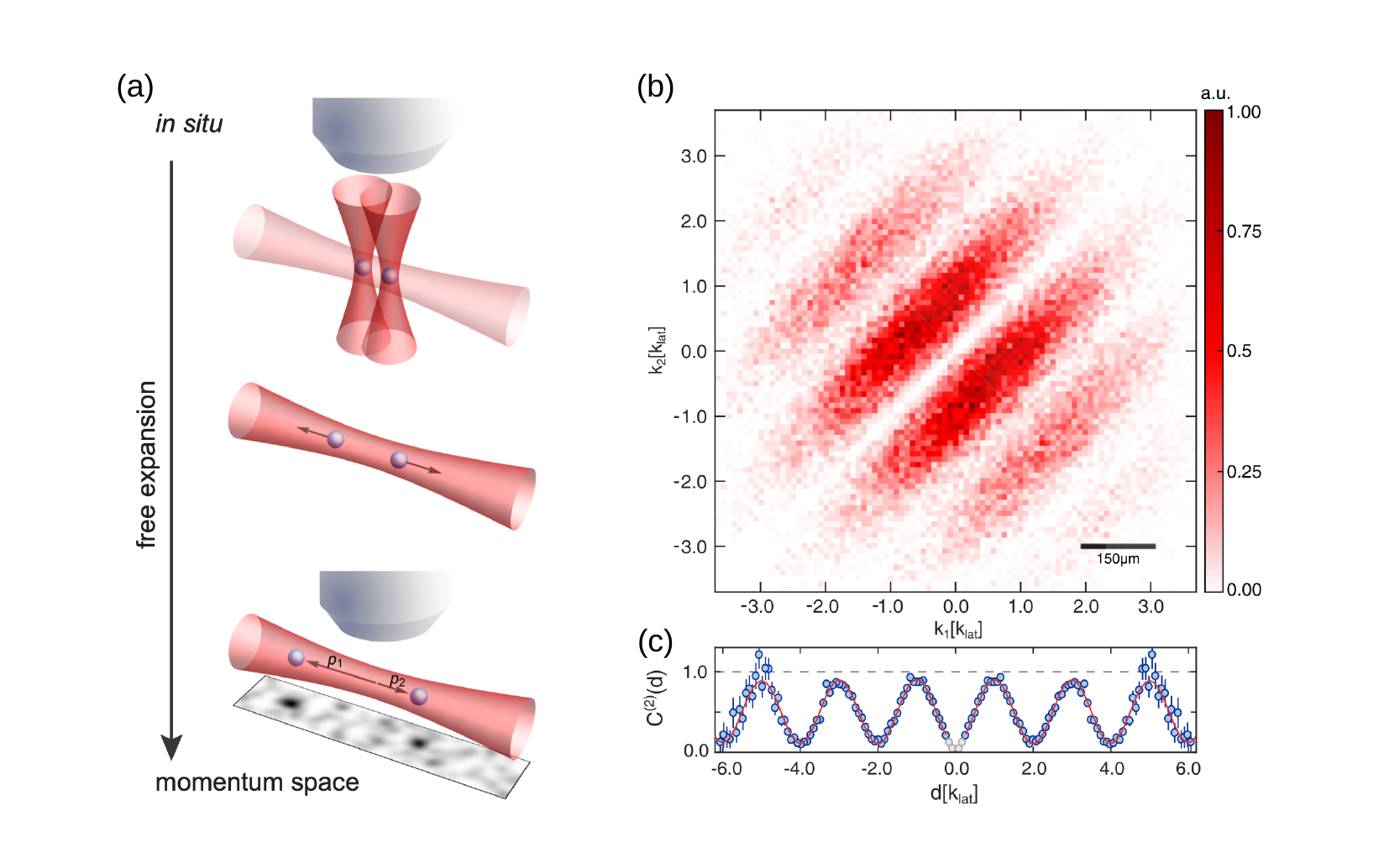}
\caption{
Ghosh and Mandel experiment for Fermions ($^6\mathrm{Li}$). 
(a) Diagram of the temporal sequence: Initially one atom is trapped in the focus of each vertical laser beam. 
When the vertical beams are turned off, the atom wave functions spread out and overlap each other. 
They are then imaged by fluorescence and their positions are recorded.
(b) Correlation of the two atoms as a function of each of their initial momenta. 
(c) Correlation as a function of the momentum difference. 
The correlation is minimal at zero momentum difference, reflecting the exclusion principle.
Figure from Ref.~\cite{Preiss_high-contrast_2019}.
}
\label{f:FermiGM}
\end{center}
\end{figure}

A related experiment was also performed using trapped fermions \cite{Preiss_high-contrast_2019}.
A diagram of the apparatus is shown in Fig.~\ref{f:FermiGM}.
Exactly one $^6\mathrm{Li}$ atom is the ground state of each of two traps consisting of two vertical laser beams and a common horizontal beam.
The vertical beams are suddenly turned off allowing the atoms to expand freely in the horizontal direction. 
After enough time has elapsed that the two wavefunctions overlap each other, 
the atoms' positions are observed by fluorescence. 
The configuration is closely analogous to that of the Ghosh and Mandel experiment shown in Fig.~\ref{f:GM-HOM}a.
The position distribution of a single atom is smooth, but fringes appear in the 2 particle correlation, $\left< n(k_1) n(k_2) \right>$, where $n(k_i)$ is the number of atoms with momentum $\hbar k_i$.
The observed positions of the atoms have been converted to momenta knowing their initial position and their time of flight.
In Fig.~\ref{f:GM-HOM}b, the correlation is shown as a 2D image illustrating the interference fringes along the line $k_1 = - k_2$.
Part c of the figure shows the fringes as a function of the variable $d= k_1 - k_2$.
The interference contrast is 79\% and one clearly sees that the fringes have a node at $d=0$ as predicted by Eq.~\eqref{e:GM-Fermi} with $d$ playing the role of $x_c-x_d$. 
Another recent experiment realized an analog of the Ghosh and Mandel experiment for fermions using two electrons originating from the double ionization of an $\mathrm H_2$ molecule 
\cite{waitz_two-particle_2016}. 

\section{Conclusion}
The source and detector configurations in the \HBT and Hong Ou Mandel experiments are frequently encountered in quantum optics and quantum information research.
We have shown that whenever the sources are ``classical" (in particular thermal or coherent states), the results have entirely classical explanations. 
Only when the sources are non-classical (our examples were number states or fermionic states) is the quantum theory necessary to account for the observations.
One might wonder how other non-classical states (a Schroedinger cat state for example) behave in these configurations. 
This could be interesting work for the future. 

\newpage
\section*{Acknowledgments}
I am grateful to my collaborators Denis Boiron, Alain Aspect and many graduate students for their input over many years.



\bibliographystyle{unsrt}
\bibliography{HBT,HOM}

\end{document}